\documentclass[12pt]{article}

\usepackage{amsmath}
\usepackage{amssymb}
\usepackage{amsfonts}
\usepackage{epsfig}
\usepackage{float}
\usepackage[a4paper]{geometry}

\usepackage{color}

\def\d{{\rm d}}
\def\i{{\rm i}}
\def\df#1#2{\frac{{\rm d}#1}{{\rm d}#2}} 
\def\pd#1#2{\frac{\partial#1}{\partial#2}} 
\def\fd#1#2{\frac{\delta#1}{\delta#2}} 

\def\HH{{\cal H}}
\def\h{h}

\def\ave#1{\left\langle#1\right\rangle}

\def\ang#1#2{\left\langle#1,#2\right\rangle}

\begin{document}

\title{Action principle for\\ nonlinear parametric quantization of gravity}

\author{Charles Wang\\
Department of Physics, Lancaster University\\
Lancaster LA1 4YB, England\\
E-mail: c.wang@lancaster.ac.uk}

\date{}

\maketitle

\abstract{The derivation of the recently proposed
nonlinear quantum evolution of gravity from an action principle
is considered in this brief note. It is shown to be possible if a set of
consistency conditions are satisfied that are analogous to the
Dirac relations for the super-Hamiltonian and momenta in classical canonical gravity.}

\vskip 7mm


\vskip 10mm

Recently the quantization of a parameterized theory in
finite dimensions with canonical coordinates $q^\mu(t), p_\mu(t)$, ($\mu=0,1,\cdots n$),
lapse function $N(t)$ with a parameter time $t$
and Hamiltonian $H(q^\mu, p_\mu, N) = N \HH(q^\mu, p_\mu)$ for some $\HH$
was considered based on a nonlinear parametric approach  \cite{wang_2003}.
A new quantization scheme was set up by choosing one of the
variables, say $\tau:=q^0$ and its conjugate momentum $\varpi:=p_0$, as classical variables that
interact with
the remaining quantized variables $q^a,\; (a=1,2,\cdots, n)$ ``semi-classically'':
\begin{equation}
\i \pd{{\psi}}{t} = \hat{H} \psi
\label{relweq}
\end{equation}
\begin{eqnarray}
\df{\tau}{t} &=& \ave{\pd{\hat{H}}{\varpi}}, \quad  \df{\varpi}{t} = - \ave{\pd{\hat{H}}{\tau}}
\label{reldott}
\end{eqnarray}
\begin{equation}
\ave{\hat{\HH}} = 0
\label{relaveH}
\end{equation}
where
$\psi = \psi(q^a,t)$ and $\ave{\hat{O}}$ denotes the expectation value of
any operator $\hat{O}$. \footnote{Units in which $c = \hbar = 16 \pi G = 1$ are adopted.}
The operators $\hat{\HH}$ and $\hat{H} = N \hat{\HH}$
are obtained by substituting $p_a \rightarrow \hat{p}_a := -\i \pd{}{q^a}$
into ${\HH}$ respectively (followed by a suitable factor ordering.)
The
presence of \eqref{reldott} and \eqref{relaveH} makes the quantum evolution {\em nonlinear}.
Like many other nonlinear quantum theories previously considered in the literature
(e.g. \cite{kibble}), the system of evolution
\eqref{relweq}, \eqref{reldott} and \eqref{relaveH}
can be derived from an action principle where the quantum and classical
variables involved are varied. To see this
first note that \eqref{relweq} may be replaced by
\begin{equation}
\i \pd{{\psi}}{t} = \hat{h} \psi
\label{relweq1}
\end{equation}
where
\begin{equation}\label{}
\hat{h} := \hat{H} - \varpi\dot{\tau}.
\end{equation}
Since $\hat{h}$ differs from $\hat{H}$ only by a function of $t$
the resulting wavefunctions are related by a time-dependent phase.
Clearly equations \eqref{relweq1}, \eqref{reldott} and \eqref{relaveH}
arise from extremizing the action
\begin{equation}\label{act}
S[\psi(q^a,t), \tau(t), \varpi(t), N(t)] := \int \Re \ang{\psi}{ \left(\i \partial_{t} - \hat{h}\right) \psi} \d t
\end{equation}
with respect to  $\psi$ and its conjugate, $\tau$, $\varpi$ and $N$.

This formulation suggests a
parametric quantization of gravity
whose formal treatment can be outlined as follows.
Start from the Dirac-ADM Hamiltonian for canonical general relativity \cite{Dirac-ADM}:
\begin{equation}\label{Dirac-ADM}
H[g_{ij}, p^{ij}, N_\mu]
:=
\int  N_\mu \HH^\mu(g_{ij}, p^{ij})\, \d^3 x
\end{equation}
in terms of the 3-metric components $g_{ij}(x)=g_{ij}(x^k,t)$ and their
conjugate momenta $p^{ij}(x)$,
lapse function $N_0(x) = N(x)$, shift functions $N_i(x)$,
super-Hamiltonian $\HH^0 = \HH$ and super-momenta $\HH^i$. ($\mu=0,1,2,3; i,j,k=1,2,3$.)
A natural way of isolating the ``true'' gravitational degrees of freedom
is to canonically transform from $g_{ij}(x), p^{ij}(x)$ to a set of four
``embedding variables'' $\vartheta^\mu(x)$ with conjugate momenta
$\varpi_\mu(x)$, ($\mu = 0,1,2,3$)
and a set of two unconstrained variables $q^{{r}}(x)$ with conjugate momenta
$p_r(x)$, (${{r}}=1,2$) \cite{embedding}.
This allows us to re-express
$H$ as $H[q^{{r}}, p_{r}, \vartheta^\mu, \varpi_\mu, N_\mu]$.
The standard interpretation of the former set is that
$\vartheta^0$ specifies time slicing whereas $\vartheta^i$  set a spatial coordinate condition.
For our discussion, the key idea is to treat $\vartheta^\mu$ as constrained classical variables
coupled to the quantized true degrees of freedom carried by $q^{{r}}$ via the
Hamiltonian operator
$\hat{H}[q^{{r}}, \hat{p}_{r}, \vartheta^\mu, \varpi_\mu, N_\mu]$
obtained from $H$ using $p_r \rightarrow \hat{p}_r := -\i \pd{}{q^r}$ with a suitable
factor ordering.

Guided by the discussions above we now construct the
action integral
\begin{equation}\label{actg}
S[\psi[q^{r}(x);t), \vartheta^\mu(x), \varpi_\mu(x),  N_\mu(x)]
:=
\int \Re \ang{\psi}{ \left(\i \partial_{t} - \hat{h}\right) \psi} \d t
\end{equation}
where
\begin{align}\label{}
\hat{h}[q^{{r}}, \varpi_{{r}}, \vartheta^\mu, \dot{\vartheta}^\mu, N_\mu]
&:= \hat{H} - \int \dot{q}^{r} p_{{r}} \, \d^3 x .
\end{align}
{\em The nonlinear quantum evolution of
gravity} is generated by varying \eqref{actg}
with respect to the state functional $\psi[q^{r}(x);t)$ and its conjugate,
and the classical variables
$ \vartheta^\mu(x)$, $ \varpi_\mu(x)$ and $N_\mu(x)$. This yields
\begin{equation}
\i \pd{{\psi}}{t} = \hat{\h} \psi
\label{relweq00}
\end{equation}
\begin{eqnarray}
\pd{\vartheta^\mu}{t} &=& \ave{\pd{\hat{H}}{\varpi_\mu}}, \quad
\pd{\varpi_\mu}{t} = - \ave{\pd{\hat{H}}{\vartheta^\mu}}
\label{reldott00}
\end{eqnarray}
\begin{equation}
\ave{\hat{\HH}^\mu} = 0.
\label{relaveH00}
\end{equation}
The structure of the above system is similar to that of \eqref{relweq1},
\eqref{reldott} and \eqref{relaveH}. However, it is necessary to
establish the consistency condition of this system. It follows from
\eqref{relweq00} and
\eqref{reldott00} that

\begin{equation}\label{}
\pd{}{t}\ave{\hat{\HH}^\mu} = \{\hat{\HH}^\mu, \hat{H}\}
\end{equation}
where\begin{equation}\label{}
\{\hat{A}, \hat{B}\}
:=
\int\left(
\ave{\fd{\hat{A}}{\vartheta^\mu(x)}}
\ave{\fd{\hat{B}}{\varpi_\mu(x)}}
-
\ave{\fd{\hat{B}}{\vartheta^\mu(x)}}
\ave{\fd{\hat{A}}{\varpi_\mu(x)}}
\right)\d^3 x
-
\i \ave{[ \hat{A}, \hat{B} ]}
\end{equation}
for any operators $\hat{A}$ and $\hat{B}$. Provided that
a set of embedding variables $\varpi_\mu$
and an operator ordering in defining $\hat{\HH}^\mu$
can be found so that
\begin{equation}\label{cc}
\{\hat{\HH}^\mu, \hat{H}\} = C^\mu_\nu\ave{\hat{\HH}^\nu}
\end{equation}
for some coefficients $C^\mu_\nu$ (as functions of spacetime points), the consistency
of the evolution system \eqref{relweq00}, \eqref{reldott00} and \eqref{relaveH00}
will be satisfied. It is worth noting that the form of \eqref{cc} is
closely analogous to that derived from
the ``Dirac algebra'' in classical canonical gravity. Further progress on
the validity of this consistency condition will be reported elsewhere.

\end{document}